\documentclass[final, conference]{IEEEtran}
\IEEEoverridecommandlockouts

\usepackage{cite}
\usepackage{amsmath,amssymb,amsfonts}
\usepackage{algorithmic}
\usepackage{graphicx}
\usepackage{textcomp}
\usepackage{xcolor}
\usepackage{siunitx}
\usepackage{balance}
\usepackage{booktabs}
\usepackage{url}
\usepackage{subcaption}
\usepackage{hyperref}
\usepackage{booktabs}
\usepackage[left=1.62cm,right=1.62cm,top=1.72cm]{geometry}
\usepackage{tabularx}
\usepackage{acronym}
\usepackage[alph]{parnotes}

\def\BibTeX{{\rm B\kern-.05em{\sc i\kern-.025em b}\kern-.08em
    T\kern-.1667em\lower.7ex\hbox{E}\kern-.125emX}}

\newcommand{\newspacing}{\baselineskip=0.97\normalbaselineskip}

\begin{document}

\title{At the Edge of the Heart: ULP FPGA-Based CNN for On-Device Cardiac Feature Extraction in Smart Health Sensors for Astronauts
\thanks{The biomedical data used in this study were collected with approval from the Ethics Committee of Hamburg University of Technology.}
\thanks{\textcopyright~2026 IEEE. Personal use of this material is permitted. Permission from IEEE must be obtained for all other uses, in any current or future media, including reprinting/republishing this material for advertising or promotional purposes, creating new collective works, for resale or redistribution to servers or lists, or reuse of any copyrighted component of this work in other works.}
}

\author{
\IEEEauthorblockN{
Kazi Mohammad Abidur Rahman\textsuperscript{1},
Davis Rakhshan\textsuperscript{1},
Philipp Lütke\textsuperscript{1},
Laura Harms\textsuperscript{3,2},
Ulf Kulau\textsuperscript{1}
}
\IEEEauthorblockA{
\textsuperscript{1}\textit{Smart Sensors Group, Hamburg University of Technology}, Hamburg, Germany\\
\textsuperscript{2}\textit{Networked Cyber-Physical Systems, Hamburg University of Technology}, Hamburg, Germany\\
\textsuperscript{3}\textit{Distributed Systems, Kiel University}, Kiel, Germany\\
kazi.rahman@tuhh.de, davis.rakhshan@tuhh.de, philipp.luetke@tuhh.de, laura.harms@cs.uni-kiel.de, ulf.kulau@tuhh.de
}
}

\newacro{WSN}[WSN]{Wireless Sensor Network}
\newacro{DVS}[DVS]{Dynamic Voltage Scaling}
\newacro{CMOS}[CMOS]{Complementary Metal-Oxide Semiconductor}
\newacro{PC}[PC]{Personal Computer}
\newacro{COTS}[COTS]{Commercial Off-The-Shelf}
\newacro{MCU}[MCU]{microcontroller unit}
\newacro{ADC}[ADC]{analog-digital converter}
\newacro{ALU}[ALU]{arithmetic logical unit}
\newacro{DPM}[DPM]{dynamic power management}
\newacro{LPL}[LPL]{low power listening}
\newacro{RSSI}[RSSI]{received signal strength indicator}
\newacro{LQI}[LQI]{link quality indicator}
\newacro{SI}[SI]{secure instance}
\newacro{LDO}[LDO]{linear dropout converter}
\newacro{I2C}[I\textsuperscript{2}C]{Inter-Integrated Circuit}
\newacro{UART}[UART]{Universal Asynchronous Receiver/Transmitter}
\newacro{GPIO}[GPIO]{General Purpose Input/Output}
\newacro{SPI}[SPI]{Serial Peripheral Interface Bus}
\newacro{UART}[UART]{Universal Asynchronous Receiver Transmitter}
\newacro{PCB}[PCB]{Printed Circuit Board}
\newacro{SI}[SI]{Secure Instance}
\newacro{IoT}[IoT]{Internet of Things}
\newacro{IC}[IC]{Integrated Circuits}
\newacro{SEU}[SEU]{single event upset}
\newacro{CPS}[CPS]{Cyber Physical Systems}
\newacro{ISP}[ISP]{In System Programmer}
\newacro{PoE}[PoE]{Power over Ethernet}
\newacro{PPU}[PPU]{Pre-Processing Unit}
\newacro{PRR}[PRR]{Packet Reception Ratio}
\newacro{INBED}[INBED]{Inexpensive Node for Bed Exit Detection}
\newacro{PLRI}[PLRI]{Peter L. Reichertz Institute for Medical Informatics}
\newacro{IBR}[IBR]{Institute for Operating Systems and Computer Networks}
\newacro{BC}[BC]{broadcast}
\newacro{UC}[UC]{unicast}
\newacro{ULP}[ULP]{ultra-low-power}
\newacro{SoC}[SoC]{System on Chip}
\newacro{MEMS}[MEMS]{Micro-Electro-Mechanical Systems}
\newacro{SW}[SW]{Software}
\newacro{HW}[HW]{Hardware}
\newacro{BCG}[BCG]{Ballistocardiography}
\newacro{ECG}[ECG]{Electrocardiography}
\newacro{FPGA}[FPGA]{Field Programmable Gate Array}
\newacro{PS}[PS]{Processing System}
\newacro{SCG}[SCG]{Seismocardiography}
\newacro{EMIO}[EMIO]{Extended Multiplexed Input/Output}
\newacro{MPSoC}[MPSoC]{Multi-Processor System on Chip}
\newacro{PTP}[PTP]{Precision Time Protocol}
\newacro{RTC}[RTC]{Real Time Clock}
\newacro{PHC}[PHC]{Physical Hardware Clock}
\newacro{MAD}[MAD]{median absolute deviation}
\newacro{TCP}[TCP]{Transmission Control Protocol}
\newacro{ADC}[ADC]{Analog-Digital-Converter}
\newacro{MMCM}[MMCM]{Mixed Mode Clock Manager}
\newacro{VCO}[VCO]{Voltage Controlled Oscillator}
\newacro{DMA}[DMA]{Direct Memory Access}
\newacro{BpA}[BpA]{Bytes per Axis}
\newacro{AB}[AB]{Address Bytes}
\newacro{BB}[BB]{Burst Bytes}
\newacro{IF}[IF]{Idle Factor}
\newacro{ILA}[ILA]{Integrated Logic Analyzer}
\newacro{PL}[PL]{Programmable Logic}
\newacro{TIE}[TIE]{Time Interval Error}
\newacro{PPSMS}[PPSMS]{Precise, Parallel, and Scalable Measurement System}
\newacro{AMBA}[AMBA]{Advanced Microcontroller Bus Architecture}
\newacro{AXI}[AXI]{Advanced eXtensible Interface Bus}
\newacro{TB}[TB]{Transmission Bytes}
\newacro{DHCP}[DHCP]{Dynamic Host Configuration Protocol}
\newacro{CS}[CS]{Chip Select}
\newacro{MISO}[MISO]{Master In Slave Out}
\newacro{MOSI}[MOSI]{Master Out Slave In}
\newacro{CPOL}[CPOL]{Clock Polarity}
\newacro{CPHA}[CPHA]{Clock Phase}
\newacro{PTT}[PTT]{Pulse Transit Time}
\newacro{MAC}[MAC]{multiply-accumulate}
\newacro{IP}[IP]{Internet Protocol}
\newacro{ISS}[ISS]{International Space Station}
\newacro{EMI}[EMI]{Electromagnetic Interference}
\newacro{SNR}[SNR]{Signal-to-Noise Ratio}
\newacro{LVDS}[LVDS]{Low Voltage Differential Signaling}
\newacro{OPAMP}[OPAMP]{Operational amplifier}
\newacro{FFT}[FFT]{Fast Fourier Transform}
\newacro{ODR}[ODR]{Output Data Rate}
\newacro{SET}[SET]{Single Event Transient}
\newacro{SEE}[SEE]{Single Event Effect}
\newacro{DAC}[DAC]{Digital Analog Converter}
\newacro{UWB}[UWB]{Ultra Wideband}
\newacro{EMC}[EMC]{Electromagnetic Compatibility}
\newacro{LDO}[LDO]{Low-Dropout Regulator}
\newacro{IMU}[IMU]{Inertial Measurement Unit}
\newacro{IR-UWB}[IR-UWB]{Impulse Radio Ultra Wideband}
\newacro{BLE}[BLE]{Bluetooth Low Energy}
\newacro{CNN}[CNN]{Convolutional Neural Network}
\newacro{LUT}[LUT]{Look-up table}
\newacro{DSP}[DSP]{digital signal processor}
\newacro{LEO}[LEO]{Low Earth Orbit}
\newacro{HR}[HR]{heart rate}
\newacro{HRV}[HRV]{heart rate variability}
\newacro{SDR}[SDR]{systole-diastole ratio}
\newacro{SVM}[SVM]{Support Vector Machine}
\newacro{KNN}[KNN]{K-Nearest Neighbors}
\newacro{QAT}[QAT]{Quantization-Aware Training}
\newacro{PE}[PE]{processing element}
\newacro{TPU}[TPU]{Tensor Processing Unit}
\newacro{NPU}[NPU]{Neural Processing Unit}
\newacro{GPU}[GPU]{Graphics Processing Unit}
\newacro{SPRAM}[SPRAM]{Single-Port RAM}
\newacro{BRAM}[BRAM]{Block RAM}
\newacro{GAP}[GAP]{global average pooling}
\newacro{FC}[FC]{fully-connected}
\newacro{ROM}[ROM]{read only memory}
\newacro{FSM}[FSM]{finite state machine}
\newacro{ECE}[ECE]{expected calibration error}

\maketitle

\begin{abstract}

The convergence of accelerating human spaceflight ambitions and critical terrestrial health monitoring demands is driving the unprecedented requirements for reliable and real-time feature extraction capabilities on extremely resource-constraint wearable health sensors.
We present a \acs{ULP} \acs{FPGA}-based solution for real-time \ac{SCG} feature classification using \acp{CNN}.
Our approach combines quantization-aware training with a systolic-array accelerator to enable efficient integer-only inference on the Lattice iCE40UP5K \acs{FPGA}, which offers an ideal platform for battery-powered deployments -- particularly in space environments -- for their power efficiency and radiation resilience.
The implementation achieves a validation accuracy of 98\% while consuming only 8.55 mW of power, completing inference in 95.5 ms with minimal hardware resources (2,861 \acsp{LUT} and 7 \acs{DSP} blocks). These results demonstrate that fully on-device \ac{SCG}-based cardiac feature extraction is feasible on resource-constrained hardware, enabling energy-efficient, autonomous health monitoring for astronauts in long-duration space missions.
\end{abstract}


\newspacing

\section{Introduction}
\label{sec:introduction}
As humanity pushes toward sustained exploration of the Moon and Mars, safeguarding astronaut health, one of the most critical priorities in space missions, demands a new generation of intelligent, wearable health monitoring systems capable of operating far beyond \ac{LEO}.
Unlike terrestrial or near-Earth deployments, deep-space missions impose severe communication constraints, where latency to ground stations can extend to minutes, rendering continuous remote supervision infeasible.
Consequently, physiological monitoring systems must achieve unprecedented levels of autonomy, reliability, and real-time decision-making to enable timely health assessment and intervention.

Building upon insights from previous cardiovascular monitoring experiments conducted aboard the \ac{ISS} \cite{kulau2022differential, drobczyk2021wireless, drobczyk2022wireless}, this work fosters the development of smart, space-grade wearable sensors specifically designed for continuous cardiac monitoring. \acf{SCG}, which captures the mechanical vibrations of the heart using inertial sensors, offers a non-invasive and compact sensing modality well suited for long-duration missions.
However, robust extraction of clinically meaningful cardiac features from \ac{SCG} signals, particularly systolic and diastolic intervals, remains challenging due to motion artifacts, physiological variability, and altered cardiac mechanics in microgravity.

Recent advances in deep learning have demonstrated that \acp{CNN} can robustly extract cardiac features from \ac{SCG} signals by learning hierarchical temporal-spectral representations directly from raw data \cite{wang2025deep, konnova2020application}.
While \acp{CNN} enable accurate systolic-diastolic segmentation and feature extraction, their computational complexity presents a significant challenge for deployment on wearable, spaceborne platforms.
Deep-space health monitoring systems must operate under extreme constraints: they must be highly miniaturized to meet spacecraft mass and volume limitations and exhibit exceptional energy efficiency to support prolonged autonomous operation.
In addition the radiation environment limits available processing components that could be used within this harsh environment.
These combined requirements preclude the use of conventional radiation-hardened components, which are typically bulky and power-hungry.

To address these constraints, this work adopts a carefully selected \ac{COTS} design strategy.
A key enabler of this approach is the Lattice iCE40UP-series \ac{ULP} \ac{FPGA}, which offers milliwatt-level power consumption alongside demonstrated resilience to radiation-induced effects \cite{rahman2024isfd}.
This \ac{ULP} \ac{FPGA} provides a unique balance between energy efficiency, flexibility, and reliability, making it well suited for continuous, real-time physiological monitoring in space.
However, deploying \ac{CNN} inference on such resource-constrained hardware requires algorithm-architecture co-design to overcome limitations in logic resources, memory capacity, and arithmetic precision.

This paper demonstrates that \ac{ULP} \acp{FPGA} using co-designed \ac{CNN} models, quantization-aware training, and a systolic-array accelerator are capable of classifying systolic and diastolic phases in wearable \ac{SCG} signals accurately and in real-time.

The outline of this paper is as follows: Section \ref{relatedWork} reviews related work, followed by short introduction to the space-grade \ac{SCG} health sensor hardware architecture in Section \ref{HW_platform}.
Section \ref{sec:data_acq} describes \ac{SCG} data acquisition and labeling, while Section \ref{sec:framework} presents the \ac{CNN} design and its optimization for resource-constrained devices.
Section \ref{sec:fpga_inference} details the \ac{ULP} \ac{FPGA} inference implementation\footnote{Implementation code available at: \url{https://github.com/Smart-Sensors-Group/NN_SysArray.git}}.
Section \ref{sec:evaluation} evaluates accuracy and performance, and Section \ref{sec:conclusion} concludes the paper.

\section{Related Works}
\label{relatedWork}
Continuous cardiovascular monitoring is essential for astronaut health, as microgravity induces cardiac deconditioning and altered hemodynamics \cite{garrett2019nasa}.
Wearable \ac{SCG} offers a non-invasive, lightweight modality for monitoring cardiac mechanical activity in space; however, reliable analysis beyond \ac{HR} and \ac{HRV} remains challenging due to motion artifacts, inter-cycle variability, sensor placement sensitivity, and microgravity-induced morphological changes \cite{Inan15, Taebi2019}. Accurate systolic-diastolic segmentation is therefore critical, as key mechanical events such as mitral and aortic valve motions occur in distinct cardiac phases and enable extraction of physiologically meaningful metrics including the \ac{SDR}, which reflects ventricular ejection-filling balance and is sensitive to myocardial relaxation, coronary perfusion, and hemodynamic load \cite{Weissler1968, Bombardini2009}.
Traditional \ac{SCG} analysis based on handcrafted features and classical classifiers such as \ac{SVM} and \ac{KNN} degrades under physiological variability, whereas recent studies show that \acp{CNN} consistently outperform these approaches by learning hierarchical temporal-spectral representations directly from raw \ac{SCG} signals \cite{wang2025deep, konnova2020application}.

Despite their effectiveness, \acp{CNN} are computationally intensive, motivating hardware acceleration for real-time, on-device inference.
Existing \ac{FPGA}-based \ac{CNN} accelerators primarily target high-end platforms such as Virtex-7, Zynq, Stratix-V, and Cyclone-V \cite{qiu2016going, zhang2018dnnbuilder}.
These devices provide extensive \ac{DSP} resources, large on-chip memories, and power budgets ranging from \SI{18}{\W} to \SI{50}{\W}~\cite{xu2023survey, wei2017automated}. While suitable for high-throughput inference, such platforms are incompatible with wearable and \ac{ULP} health monitoring systems.
In contrast, resource-constrained \acp{FPGA} such as the Lattice iCE40 family operate in the milliwatt range but offer limited logic resources, minimal \ac{DSP} support, and small memory capacities, rendering conventional \ac{CNN} mapping strategies impractical.

For \ac{CNN} inference on ultra-constrained devices, hardware capabilities and on-chip storage must be carefully considered, as memory is primarily allocated to model parameters, including weights and biases.
Floating-point inference is generally infeasible on \ac{ULP} \acp{FPGA}; therefore, quantization is essential.
While post-training quantization can reduce model size, \ac{QAT} explicitly models reduced-precision arithmetic during training, enabling accurate integer-only inference at low bit widths.
This approach significantly reduces memory footprint and aligns computation with resource-constrained hardware datapaths~\cite{jacob2018quantization, khudia2021fbgemm}, making it particularly suitable for \ac{ULP} \ac{FPGA}-based \ac{CNN} deployment.

Systolic array architectures, introduced by H.T. Kung~\cite{kung1982systolic}, offer an efficient solution for \ac{CNN} inference on resource-constrained \acp{FPGA}.
Composed of two-dimensional grids of \acp{PE} with local registers, they enable pipelined data movement and efficient data reuse~\cite{samajdar2020systematic, xu2023survey}.
By propagating intermediate results between adjacent \acp{PE}, systolic arrays reduce on-chip and off-chip memory accesses, lowering energy consumption, memory bandwidth requirements, and I/O bottlenecks while maintaining balanced workloads~\cite{genc2021gemmini, xu2023survey}.

Hence, systolic arrays are well matched to \acp{CNN} because convolution operations rely heavily on matrix and tensor multiplications that map naturally onto their pipelined structure~\cite{xu2023survey}.
The predictable computation patterns of \acp{CNN} further mitigate issues related to irregular parallelism.
Consequently, systolic arrays form the core of modern \acp{TPU}, \acp{NPU}, and \acp{GPU} and are widely used in signal and image processing applications~\cite{jouppi2017datacenter}.
These characteristics enable real-time, on-device \ac{SCG}-based cardiac feature extraction on \ac{ULP} \ac{FPGA} platforms.

\section{Space-grade \acs{SCG} Health Sensor}\label{HW_platform}

\begin{figure}
    \centering
    \includegraphics[width=0.95\linewidth]{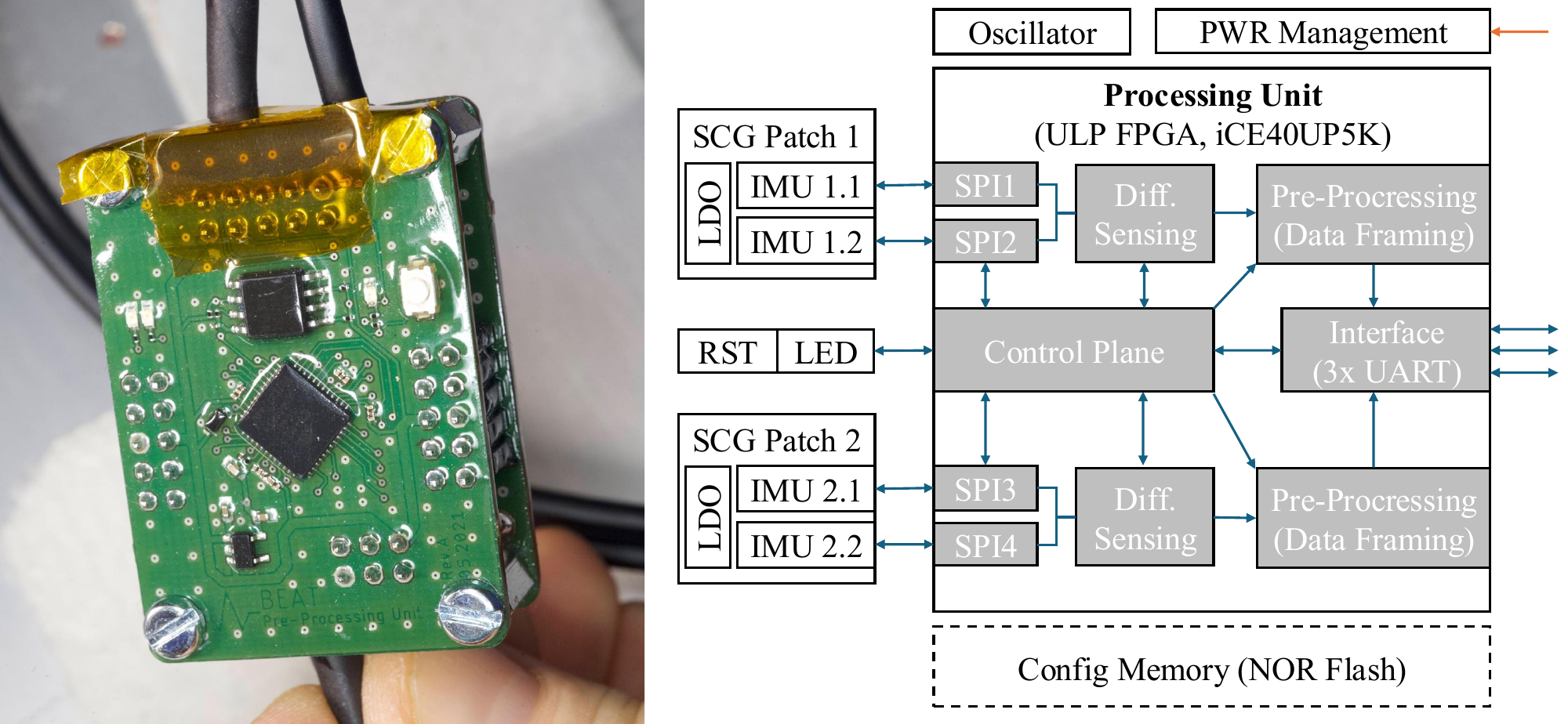}
    \caption{The processing unit of the wearable health sensor used by astronauts during ESA's Cosmic Kiss mission on the International Space Station (ISS)~\cite{kulau2022differential} }
    \label{fig:sensor}
\end{figure}

Before presenting the proposed \ac{CNN} model and inference architecture, we introduce the underlying hardware platform and its suitability for \ac{ULP} wearable health sensing in extreme and radiation-exposed environments.
For space applications, \acp{FPGA} are typically preferred over \acp{MCU} due to their higher radiation tolerance, computational throughput, deterministic timing, and fine-grained architectural controllability.
The Lattice iCE40UP5K \ac{ULP} \ac{FPGA} \cite{iCE40_DS} aligns well with these requirements and has demonstrated suitability through prior flight heritage \cite{kulau2022differential} and radiation characterization studies \cite{rahman2024isfd}.

Figure~\ref{fig:sensor} illustrates the iCE40UP5K-based processing unit used in ESA’s \emph{Cosmic Kiss} mission on the \ac{ISS} \cite{drobczyk2022wireless}, which also serves as the base sensor platform for this work.
This platform enables two-channel \ac{SCG} acquisition using high-precision \acp{IMU}.
Each \ac{SCG} channel (Patch 1, Patch 2) uses differential sensing~\cite{kulau2022differential} to facilitate ultra-low-noise signal acquisition.
The system samples at \SI{1}{\kHz} with a precision of $61\mu$\,g, and \ac{FPGA}-based acquisition ensures precise clock synchronization and cycle-accurate inter-channel alignment.

From a \ac{CNN} acceleration perspective, the iCE40UP5K is highly resource constrained but provides a set of ultra-low-power hardware primitives, integrating \SI{128}{\kilo\byte} of \ac{SPRAM}, organized as four \SI{32}{\kilo\byte} blocks (16K$\times$16-bit), for storing convolution weights and intermediate activation buffers, alongside 30 \ac{BRAM} instances totaling \SI{4}{\kilo\byte} that support dual-port access and synthesis-time initialization and are therefore used for precision-critical parameters such as biases.
Computation is supported by eight embedded 16$\times$16-bit \ac{DSP} blocks capable of single-cycle \ac{MAC} operations, which can be organized into a systolic \ac{MAC} structure and time-multiplexed across kernel taps and channel dimensions to efficiently execute fixed-point convolutional and fully connected layers under tight resource constraints.

\section{Data Acquisition and Labelling}
\label{sec:data_acq}
For data acquisition, we used the ground reference system of the space-grade \ac{SCG} health sensor described in Section~\ref{HW_platform} and extended this platform with a single-lead \ac{ECG} for reference purposes~\cite{albrecht2024identifying}.

For each experiment, an \ac{SCG} Sensor Patch was placed on the subject’s chest, in particular on the sternum, which is well suited regardless of the subject's gender.
Both \ac{SCG} and \ac{ECG} data were collected under ethical approval from six subjects (age range 13-40 years), including four male and two female participants.
Each subject was recorded continuously for 60 minutes.
To increase signal variability and capture physiological changes due to higher heart and breathing rates, subjects performed a brief running activity after every 15 minutes of recording.

During labeling, Systolic and Diastolic cardiac phases were annotated using a semi-automated labeling pipeline.
Initial pre-labels were generated using the wavelet-driven automated \ac{SCG} Systolic/Diastolic window extraction approach described in~\cite{rahman2025wavelet}.
These predictions were imported into Label Studio~\cite{LabelStudio} and subsequently refined through manual correction to ensure high labeling accuracy.

\section{Neural Network Design}
\label{sec:framework}

\begin{figure}[]
    \centering
    \includegraphics[width=1\linewidth,height=.2\textheight,keepaspectratio]{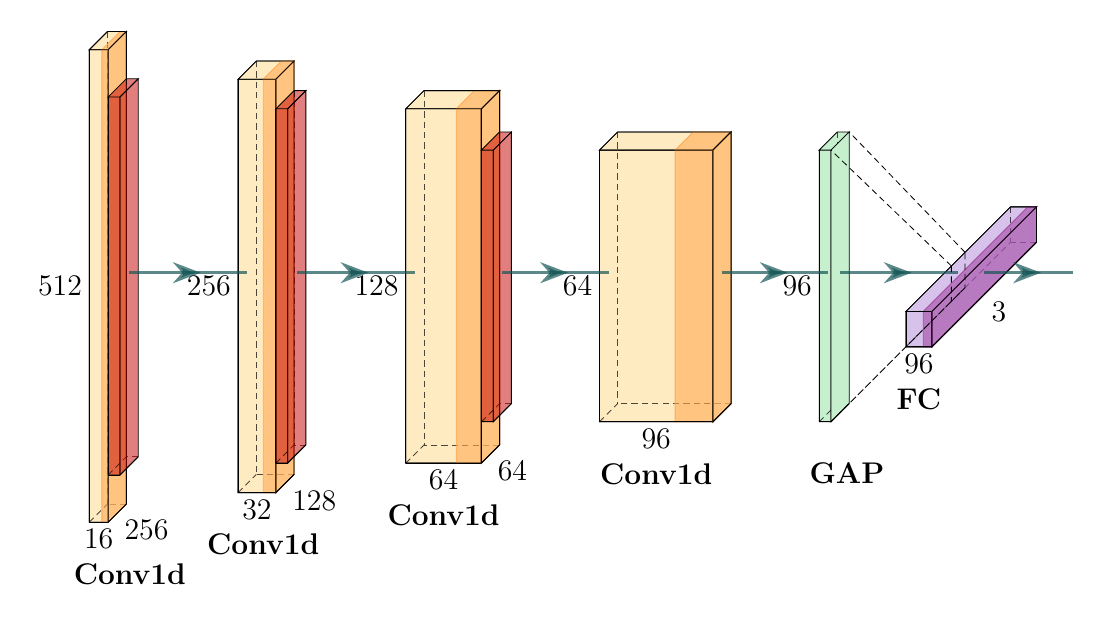}
    \caption{Neural network architecture for \ac{SCG} feature extraction. Yellow: 1D convolution, red: batch normalization, green: \acf{GAP}, violet: \acf{FC}.}
    \label{fig:SCG_NN_Architecture}
\end{figure}

To classify the cardiac phases, we build a neural network composed of a sequence of one-dimensional convolutional blocks that extract hierarchical temporal features from fixed-length \ac{SCG} signal windows. Each block consists of a 1D convolution followed by batch normalization, a ReLU activation, and max pooling.
This structure enables progressive abstraction of temporal patterns while reducing the temporal resolution of the signal.

We chose a one-dimensional convolutional neural network for its ability to learn temporal representations directly from raw signals while remaining compact and hardware-efficient. The proposed model operates on fixed-length \ac{SCG} windows and is designed to balance classification performance with the power, memory, and latency constraints required for on-device processing in smart health sensors.

The input to the network is a single-channel \ac{SCG} signal window, which is classified into one of three classes: (1) \textit{Systolic}, (2) \textit{Diastolic}, and (3) \textit{Background}.
The input signal is normalized using per-window z-score normalization to reduce amplitude variations across recordings.
During training, the target label corresponds to the class of the sample located at the center of the input window.

We illustrate the network architecture in Figure~\ref{fig:SCG_NN_Architecture}, which shows the layer-wise structure and corresponding dimensionality changes.

All convolutional layers use a kernel size of $9$, which provides sufficient temporal context to capture characteristic cardiac signal patterns while maintaining computational efficiency and a limited parameter budget.
A padding of $4$ is applied to preserve temporal resolution across these layers.
In the final convolutional layer, the kernel size is reduced to $5$ with a padding of $2$, reflecting the increased level of abstraction in deeper layers.
Three convolutional blocks are stacked with channel dimensions $16 \rightarrow 32 \rightarrow 64$, followed by a final convolutional layer of size $96$ and global average pooling.

The model is trained using a weighted cross-entropy loss to address class imbalance between background and event samples.
The loss weights emphasize systolic and diastolic events relative to background. Optimization is performed using the AdamW optimizer~\cite{loshchilov2019decoupledweightdecayregularization} with weight decay for regularization, and dropout is applied in the classification head to reduce overfitting.

The pooled features are processed by a lightweight fully connected classification head that outputs class probabilities for the three target classes.

\subsection{Optimization}
\label{subsec:optimization}

Since the target application focuses on near-sensor inference, \acf{QAT} is employed to reduce the model weights and activations to 8-bit integer precision while preserving most of the classification accuracy.
Compared to post-training quantization, \ac{QAT} generally results in a lower accuracy degradation at the cost of retraining the network.
Since the trained model is deployed on the target device exclusively for inference, the additional training cost is negligible.
During inference, the intermediate results of the multiply-accumulate operations are accumulated in higher precision and must be rescaled to match the fixed 8-bit activation format.
This re-quantization step is necessary because applying the scaling multiplier to the 32-bit accumulation produces a 64-bit intermediate value, which must then be reduced back to 8-bit.
This precision was chosen to meet strict memory constraints, since higher precision would exceed the limit or require removing model structures, leading to reduced accuracy.

\begin{table}[t]
\centering
\caption{Confusion matrix for the FP32 model on the test set. Rows denote ground truth, columns denote predictions.}
\label{tab:confusion_fp32}
\begin{tabularx}{\linewidth}{lXXX}
\toprule
\textbf{True $\backslash$ Pred} & \textbf{Background} & \textbf{Systolic} & \textbf{Diastolic} \\
\midrule
Background & 9469 & 39  & 383 \\
Systolic   & 55   & 9914 & 125 \\
Diastolic  & 44   & 45  & 9926 \\
\bottomrule
\end{tabularx}
\end{table}

\autoref{tab:confusion_fp32} presents the confusion matrix of the FP32 model evaluated on the test set. The results indicate strong class separability, with the majority of misclassifications occurring between background and event classes, while confusion between systolic and diastolic events remains limited. This behavior is consistent with the temporal proximity of cardiac events and validates the robustness of the learned feature representations.

\begin{table}[t]
\centering
\caption{Classification performance and model size comparison between FP32 and INT8 (QAT) models.}
\label{tab:fp32_vs_int8}
\begin{tabularx}{\linewidth}{lXX}
\toprule
\textbf{Metric} & \textbf{FP32} & \textbf{INT8 (\acs{QAT})} \\
\midrule
Validation accuracy [\%] & 98.68 & 98.32 \\
Test accuracy [\%]       & 97.70 & 97.68     \\
Model size [KB]          & 226.5  & 68.2   \\
\bottomrule
\end{tabularx}
\end{table}

As shown in \autoref{tab:fp32_vs_int8}, quantization results in only a marginal degradation in classification performance, with a decrease of 0.36 percentage points in validation accuracy and 0.02 percentage points in test accuracy, while reducing the model size by approximately \SI{70}{\percent}.

At inference time, several operations can be fused to further improve efficiency.
Since batch normalization parameters are fixed after training, the batch normalization operation can be absorbed into the preceding convolution by folding the scale and bias terms into the convolution weights and bias.
Additionally, the ReLU activation is implicitly fused during quantized inference, as it is performed through clamping the re-quantized outputs to the non-negative range of 8-bit integers.
Furthermore, the multiplicators used to scale the intermediate values at re-quantization can be fused in one multiplicator for inference.

\section{ULP FPGA Inference} \label{sec:fpga_inference}

\begin{figure}
    \centering
    \includegraphics[width=1\linewidth]{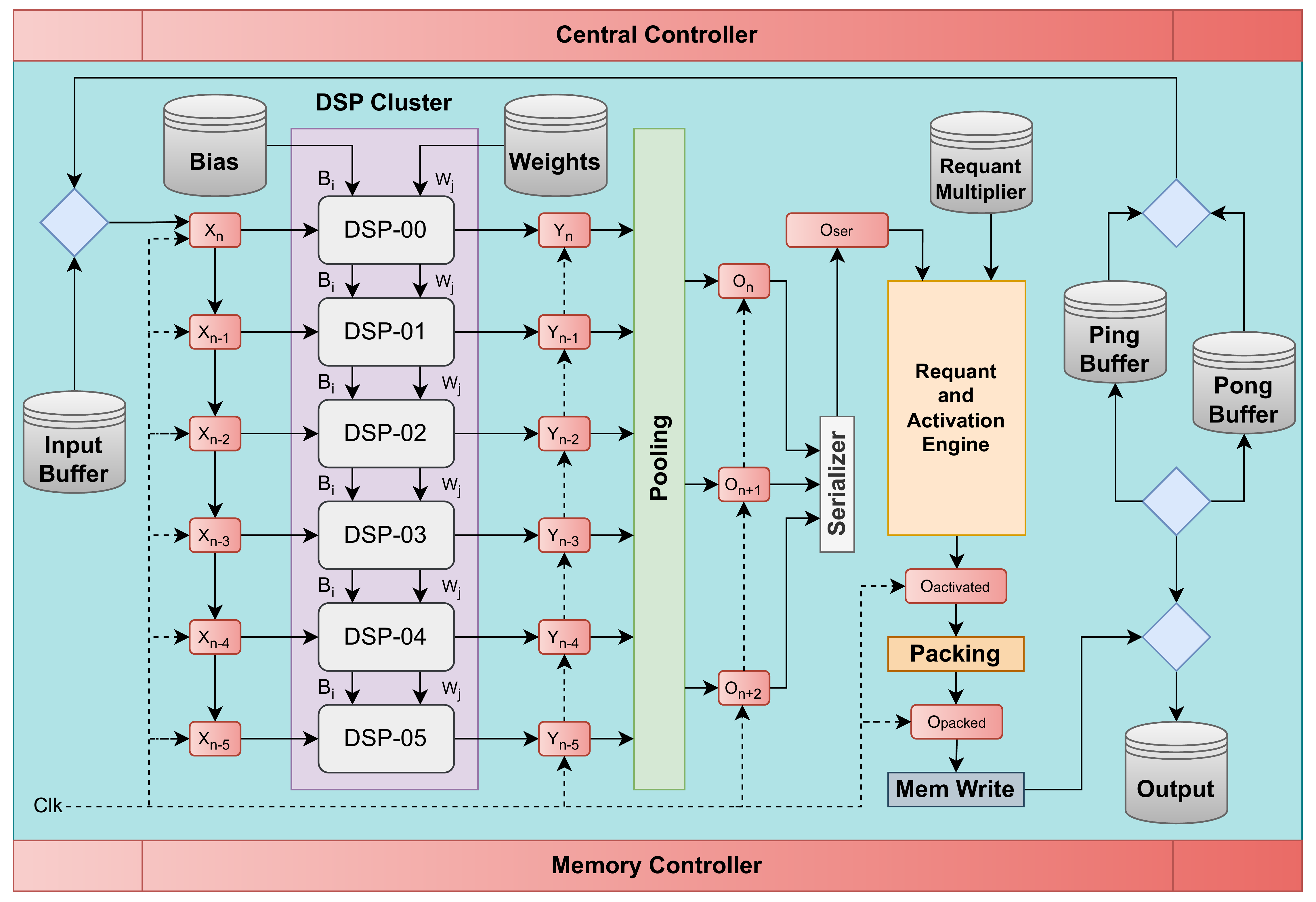}
    \caption{Proposed FPGA inference architecture. Components: Memory Subsystem (Weight Memory, Input Buffer, Ping-Pong Buffers, Bias/Scale ROMs), Compute Subsystem (DSP Systolic Cluster, Pooling Unit, Requantization \& Activation Engine, Result Packer), Control Subsystem.}
    \label{fig:fpgaInfArch}
\end{figure}

This section describes the implementation of a dedicated \ac{CNN} accelerator for one-dimensional convolutions in biomedical signals on \ac{ULP} \ac{FPGA} (iCE40UP5K), organized around a systolic-array dataflow for efficient, low-latency computation as illustrated in Figure~\ref{fig:fpgaInfArch}.

\subsection{Memory Subsystem}
\label{sec:mem_subsystem}
The proposed architecture exploits the iCE40UP5K’s heterogeneous memory, combining \ac{SPRAM} for high-density storage and dual-port \ac{BRAM} for parallel access.
The memory subsystem is organized hierarchically into four functional blocks-Weight Memory, Input Buffer, Ping-Pong Buffers, and \acsp{ROM} for biases and scaling factors-maximizing bandwidth while avoiding compute-control conflicts.

The Weight Memory stores quantized INT8 model parameters and is implemented using two cascaded 16K×16-bit \ac{SPRAM} primitives, forming a continuous 32K-word (64 KB) address space.
An address-decoding scheme uses the most significant bit (MSB) of the 15-bit address bus to select between the Lower Bank (\texttt{0x0000-0x3FFF}) and Upper Bank (\texttt{0x4000-0x7FFF}).
Two 8-bit weights are packed per 16-bit word, doubling effective read bandwidth and enabling a single fetch cycle to supply operands for two sequential multiply-accumulate operations in the compute cluster.
Weights can be dynamically loaded at runtime via the \ac{UART} interface using the packet-based loading mechanism described in Section~\ref{fsm}.

Bias and Scale \acsp{ROM} store precision-critical constants for inference.
Bias values for up to 512 output channels are held in four \verb|SB_RAM40_4K| block RAMs, each storing 32-bit values.
Scale multipliers are implemented using D flip-flops (DFFs), with one 32-bit scale value per layer.
Both Bias and Scale \acsp{ROM} are statically initialized during synthesis, ensuring immediate availability at system boot without external configuration or runtime loading.

The Input Buffer accommodates incoming 1D time-series input vectors using two \verb|SB_RAM40_4K| blocks reconfigured in native x16 mode, forming a dual-bank system with \SI{1}{\kilo\byte} total capacity (2 × 256 × 16-bit).
This allows packing of two 8-bit inputs per address, while a lightweight Read Mux in the controller supports byte-level access.

Finally, Ping-Pong Buffers manage inter-layer data transfer as zero-copy intermediate scratchpads.
Two 16k × 16-bit \ac{SPRAM} blocks operate as 'Ping' and 'Pong' buffers in a double-buffering scheme: Layer $N$ reads execution data from the 'Ping' buffer while the preceding Layer $N-1$ (or the Requantizer) simultaneously writes results into the 'Pong' buffer.
A global toggle signal swaps buffer ownership upon layer completion, eliminating memcpy overhead and ensuring continuous pipeline operation.

\subsection{Compute Subsystem}
\label{syscomp}

\subsubsection{DSP Systolic Cluster}
\label{sec:dsp_cluster}
The core of the accelerator is a one-dimensional systolic array consisting of six iCE40 \verb|SB_MAC16| \ac{DSP} blocks arranged in a pipelined sequence.
In this design, the input feature map ($X$) enters the first \ac{DSP} unit and propagates through the array via six pipeline registers (\verb|X[0..5]|), establishing a systolic data flow that reduces register fanout and balances routing delays across the \ac{FPGA} fabric.
In contrast, the filter weights and biases are simultaneously broadcast to all units.
This architecture allows the cluster to perform six \ac{MAC} operations per clock cycle, achieving a theoretical peak throughput of 144 million \acp{MAC} (MMACs) per second (at \SI{24}{\MHz}).

The architecture is layer-agnostic and supports variable kernel sizes ($K$) as well as input and output channel depths ($C_{in}, C_{out}$) through runtime configuration registers.
For scaling, large layers are time-multiplexed across the fixed-size array.
The control \ac{FSM} iterates over the kernel weights ($k=0..K$), input($c_in=0..C_{in}$) and output ($c_out=0..C_{out}$) channels, accumulating partial sums in the \ac{DSP}'s 32-bit registers before producing the final output.
On fully connected layers, the final classification layer (L5) is implemented as a special case of a $1 \times 1$ convolution.
Because the preceding \acf{GAP} reduces the spatial dimension to one, the fully connected operation is reduced to a vector-matrix multiplication.
The systolic array executes this operation efficiently, eliminating the need for a separate dense hardware unit.
The current implementation is strictly optimized for stride-one convolutions. This approach simplifies the systolic dataflow, as the pipeline registers shift by one position per cycle, precisely matching the sliding-window requirement. Supporting stride values greater than one would necessitate complex skip-connection routing or stall cycles to align the data.

\subsubsection{Pooling Unit}
Positioned after the \ac{DSP} cluster, the Pooling Unit performs dimensionality reduction with layer-specific behaviour.
For standard convolutional layers (L0-L2), it applies $2 \times 1$ Max Pooling by comparing adjacent systolic outputs (i.e. $Y_{n}$ and $Y_{n-1}$) to retain dominant features. In the final CNN layer (L3), it applies \ac{GAP}, where the input data is first scaled using arithmetic right shift (as input feature depth is fixed 64) and acts as a persistent accumulator, summing the entire feature map channel into a single 32-bit value. Once the entire input channel is processed, it flushes the accumulator.
For the fully connected layer (L4), the Bypass Mode disables all reduction logic, allowing the raw accumulation result from the systolic array to flow directly to the Requantizer. This is essential for the final classification operation, where $1 \times 1$ convolution results must be preserved exactly.

\subsubsection{Requantization and Activation Engine}
\label{sec:requant}
As discussed in Section~\ref{subsec:optimization}, for \ac{QAT} models, the pooled output (32-bit) must be scaled with a multiplier (32-bit) to retain precision before activation, producing a 64-bit intermediate value.
However, iCE40 \verb|SB_MAC16| \ac{DSP} blocks natively support only $16 \times 16$ multiplication.
To enable $32 \times 32$ signed multiplication, the proposed architecture employs a dedicated \texttt{mul64signed} unit based on a sum-of-products decomposition.
Each 32-bit operand is split into 16-bit upper ($H$) and lower ($L$) halves, and the full 64-bit product is accumulated over four clock cycles, producing one result every four cycles.

Following multiplication, the result is accumulated with a zero scale and passed through an activation unit.
For convolution layers (L0-L3), a ReLU-style activation is implemented via unsigned saturation to the range $[0, 255]$, which re-quantizes the output to 8-bit.
For the final fully connected layer (L4), the activation stage is bypassed, allowing signed 32-bit rounded logits to be forwarded directly to the host processor.

\subsubsection{Result Packer}
\label{sec:result_packer}
The final stage is the Result Packer, which adapts the 8-bit compute stream to the 16-bit memory interface. Incoming bytes are buffered in a temporary register
and two consecutive results are concatenated into a single 16-bit word.
This packing enables efficient utilization of the single-port \acp{SPRAM} (ping-pong buffers), reducing write operations by half and maximizing available memory bandwidth for subsequent layer readout.

\subsection{Control Subsystem}
\label{fsm}

\begin{figure}
    \centering
    \includegraphics[width=1\linewidth]{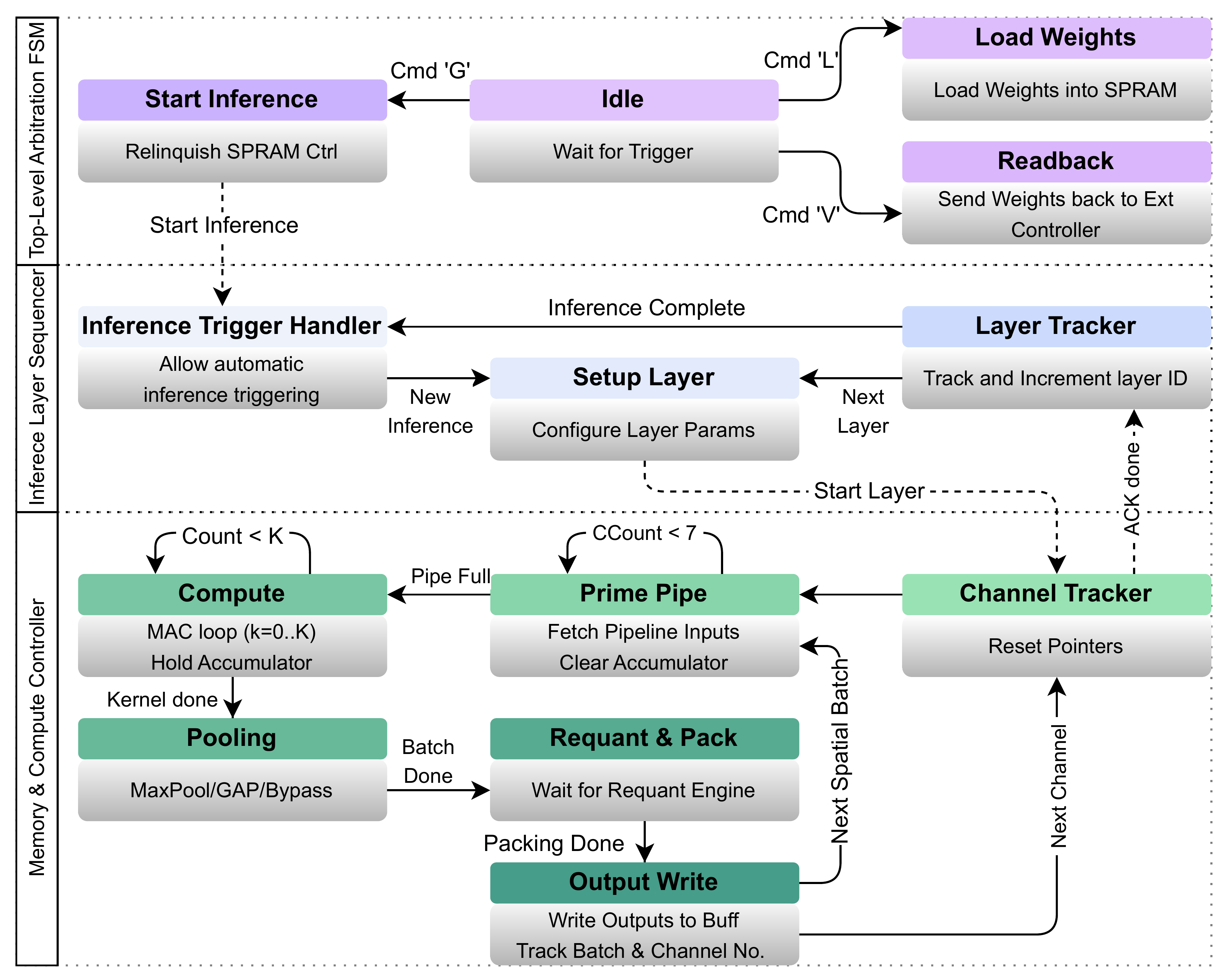}
    \caption{Top-level arbitration and control flow. The controller decodes \ac{UART} commands to switch between weight loading, memory verification, and inference modes. During loading and verification, it arbitrates exclusive access to the Weight Memory, while a run command transfers control to the inference layer sequencer for autonomous \ac{CNN} execution.}
    \label{fig:subsysCtrl}
\end{figure}

\begin{table*}[]
\caption{Layer-wise Cycle Breakdown, Compute Efficiency, and Primary Bottlenecks}
\label{tab:network_arch}
\centering
\renewcommand{\arraystretch}{1.15}
\setlength{\tabcolsep}{6pt}
\setlength{\tabcolsep}{4.2pt}
\begin{tabularx}{\linewidth}{c c c c c c c c c c c c}
\toprule
\textbf{Layer} &
\textbf{Type} &
\textbf{In (depth)} &
\textbf{Out (depth)} &
\textbf{$C_{\text{in}}\!\rightarrow\!C_{\text{out}}$} &
\textbf{$K$} &
\textbf{Prime} &
\textbf{Compute} &
\textbf{Requant / Pack} &
\textbf{Array Eff.} &
\textbf{Sys. Eff.} &
\textbf{Primary Bottleneck} \\
\midrule
L0 & Conv1D & 512 & 256 & $1 \rightarrow 16$   & 9 & 9,632   & 12,384  & 24,768  & 56\%  & 26\%  & Requantization \\
L1 & Conv1D & 256 & 128 & $16 \rightarrow 32$  & 9 & 154,112 & 198,144 & 24,768  & 56\%  & 52\%  & Balanced \\
L2 & Conv1D & 128 & 64  & $32 \rightarrow 64$  & 9 & 315,392 & 405,504 & 25,344  & 56\%  & 53\%  & Balanced \\
L3 & Conv1D & 64  & 1  & $64 \rightarrow 128$ & 5 & 630,784 & 450,560 & 768     & 41\%  & 41\%  & Priming \\
L4 & FC     & 1   & 1   & $128 \rightarrow 3$  & 1 & 2,688   & 384     & 18      & 12.5\% & 12\%  & Priming \\
\midrule
\textbf{Total} & & & & & & \textbf{1,112,608} & \textbf{1,066,976} & \textbf{75,666} & - & - & - \\
\bottomrule
\end{tabularx}
\end{table*}

The control subsystem, shown in Figure~\ref{fig:subsysCtrl}, is implemented as a hierarchical \ac{FSM} that partitions system control across three levels: a top-level arbitration controller, a layer-level inference sequencer, and a memory-compute controller.
This hierarchy decouples external communication and system orchestration from layer scheduling and arithmetic dataflow, allowing each component to be optimized independently. As a result, the arithmetic core remains agnostic to the user interface protocol while maintaining deterministic execution and low control overhead. At the highest level, the arbitration controller manages system operation and \ac{UART}-based interaction, coordinating weight loading, verification, and inference by dynamically arbitrating access to shared memory and compute resources.

Once execution is initiated, control is transferred to the \textit{Layer Sequencer}, a dedicated \ac{FSM} that encodes the neural network topology directly in hardware.
Rather than relying on an instruction fetch mechanism, the sequencer traverses the network layers using a fixed \ac{FSM} schedule, eliminating instruction memory overhead.
For each layer, it programs the memory and compute controller with layer-specific parameters, asserts a start signal to launch computation, and waits for a completion flag. Upon layer completion, the sequencer automatically advances to the next stage and reconfigures the controller as needed, enabling fully autonomous end-to-end inference once triggered.

At the lowest level of the hierarchy, the \textit{Memory and Compute Controller} governs fine-grained data movement and computation for a single layer.
It implements a four-level nested loop structure in hardware to maximize data reuse and sustain high utilization of the systolic compute array.
Output channels are processed in an outer loop, while the input signal is consumed in fixed six-element spatial batches. Each batch begins with a prime phase that pre-fetches input samples and fills the \ac{DSP} pipeline, hiding cascade latency, as detailed in Figure~\ref{fig:subsysCtrl}.
This is followed by a compute phase that iterates exclusively over the kernel depth, broadcasting weights and accumulating partial sums across all output channels. Upon completion of the compute phase, pooling, re-quantization, and result packing are applied, after which the controller advances to the next spatial batch. Valid outputs are packed into 16-bit words for efficient memory writes.

\subsection{Computational Cycle Analysis}
\label{sec:comp_cycle_analysis}
To evaluate the performance of the accelerator, we analyze the total cycle count per layer and identify the primary sources of overhead.
The execution time of a layer can be decomposed into three major components:

\begin{small}
\begin{equation}
T_{\text{Total}} = T_{\text{Prime}} + T_{\text{Compute}} + T_{\text{Requant}},
\end{equation}
\end{small}

where $T_{\text{Prime}}$ represents the pipeline priming overhead incurred when switching between input channels, $T_{\text{Compute}}$ corresponds to the convolution computation, and $T_{\text{Requant}}$ accounts for the cycles spent on re-quantization and packing. Specifically, the timing contributions are given by:

\begin{small}
\begin{equation}
\begin{cases}
T_{\text{Prime}} = C_{\text{out}} \cdot N_{\text{batches}} \cdot C_{\text{in}} \cdot 7 \\
T_{\text{Compute}} = C_{\text{out}} \cdot N_{\text{batches}} \cdot C_{\text{in}} \cdot K \\
T_{\text{Requant}} = N_{\text{outputs}} \cdot 9
\end{cases}
\end{equation}
\end{small}

where $C_{\text{in}}$ and $C_{\text{out}}$ denote the number of input and output channels, $K$ is the kernel size, $N_{\text{batches}} = \lceil W_{\text{in}}/6 \rceil$ is the number of spatial batches, and $N_{\text{outputs}}$ is the total number of outputs per layer. The 7-cycle term in $T_{\text{Prime}}$ captures the fixed latency to fill the systolic pipeline, while the 9-cycle term in $T_{\text{Requant}}$ includes six cycles for the serial multiplier and three cycles for FSM overhead and packing.

We define two efficiency metrics to quantify hardware utilization.
The array efficiency measures the fraction of cycles performing arithmetic within the systolic array, ignoring re-quantization and packing:

\begin{small}
\begin{equation}
\eta_{\text{array}} = \frac{T_{\text{Compute}}}{T_{\text{Prime}} + T_{\text{Compute}}} = \frac{K}{K + 7}
\end{equation}
\end{small}

whereas the system efficiency captures useful compute relative to the total execution time including re-quantization:

\begin{small}
\begin{equation}
\eta_{\text{sys}} = \frac{T_{\text{Compute}}}{T_{\text{Total}}}
\end{equation}
\end{small}

A layer-wise analysis reveals that early convolutional layers with large kernels achieve moderate array efficiency (about \SI{56}{\percent}) but exhibit varying system efficiency due to the relative cost of re-quantization.
For instance, Layer 0, with a 1D convolution producing 4,096 outputs, is dominated by the re-quantization stage, resulting in a system efficiency of approximately \SI{26}{\percent}, despite a pipeline utilization of \SI{56}{\percent}.
In subsequent layers (L1 and L2), the computational workload increases and the re-quantization cost becomes negligible, leading to higher system efficiency (about \SI{52}{\percent} to \SI{53}{\percent}).
Later layers, including L3 with a smaller kernel and wider channels, are dominated by pipeline priming, while the fully connected Layer L4 represents the worst-case scenario for the array, where $K=1$ limits pipeline utilization to only \SI{12.5}{\percent}, yielding similarly low system efficiency.

Table~\ref{tab:network_arch} summarizes the cycle breakdown and efficiency per layer.
The analysis highlights two primary performance bottlenecks: output-bound layers (e.g., L0), limited by the serial multiplier in the re-quantization stage, and latency-bound layers (e.g., L3 and L4), limited by pipeline priming in the systolic array.
Early convolutional layers are compute-efficient, while deeper and pointwise layers are increasingly dominated by pipeline overhead, emphasizing the impact of kernel size and channel width on array utilization and overall system performance.

Overall, the network requires approximately 2.26M cycles per inference, corresponding to about \SI{94}{\ms} at a \SI{24}{\MHz} clock frequency, yielding an effective throughput of $\sim$10.6 FPS.
This dual-bottleneck behavior provides clear guidance for optimization: either increasing the parallelism in the re-quantization stage to mitigate output-bound layers or reducing pipeline depth and kernel imbalance to improve efficiency in latency-bound layers.

\section{Evaluation}
\label{sec:evaluation}

This section presents a comprehensive evaluation of the proposed system, covering both model-level classification performance and hardware-level inference efficiency.
The analysis first examines the accuracy, robustness, and calibration of the CNN on \ac{SCG} data using qualitative examples and quantitative window-level metrics.
Inference performance of the \ac{FPGA} implementation is then evaluated with respect to resource utilization, latency, power consumption, and energy efficiency, with comparisons to a low-power \ac{MCU} baseline.

\subsection{Classification Performance}
\label{sec:classification_performance}

\begin{figure}[t]
  \centering

  \includegraphics[width=\linewidth]{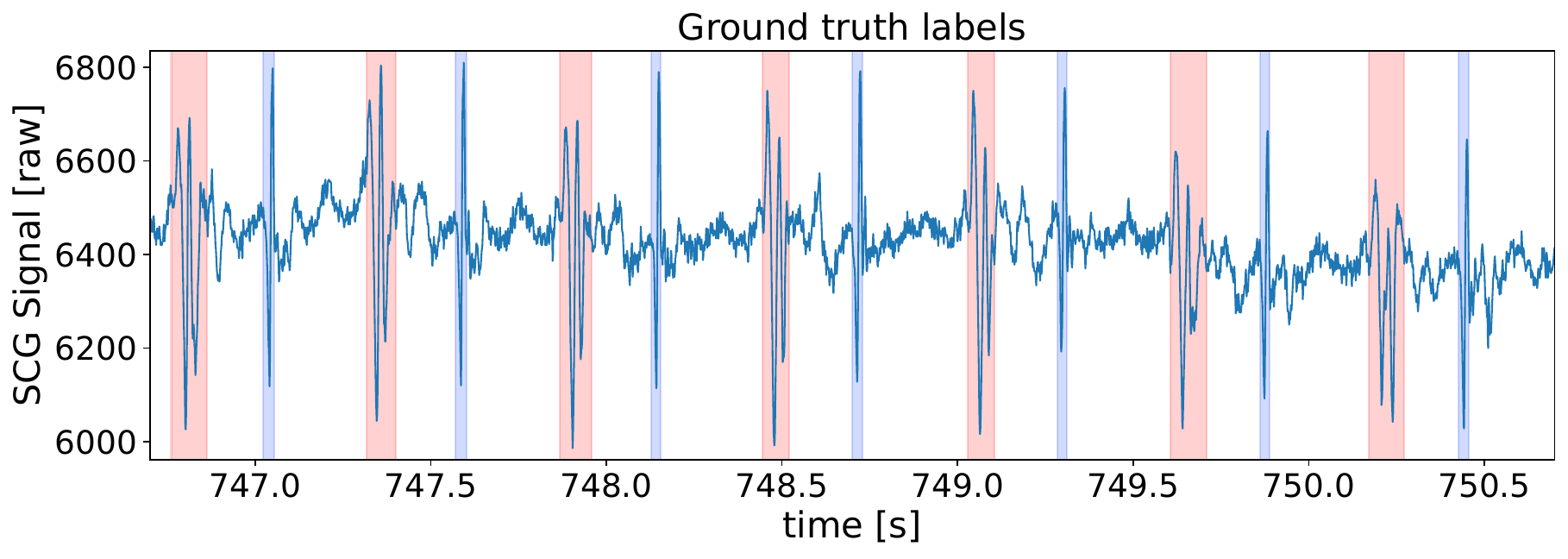}

  \vspace{0.2em}

  \includegraphics[width=\linewidth]{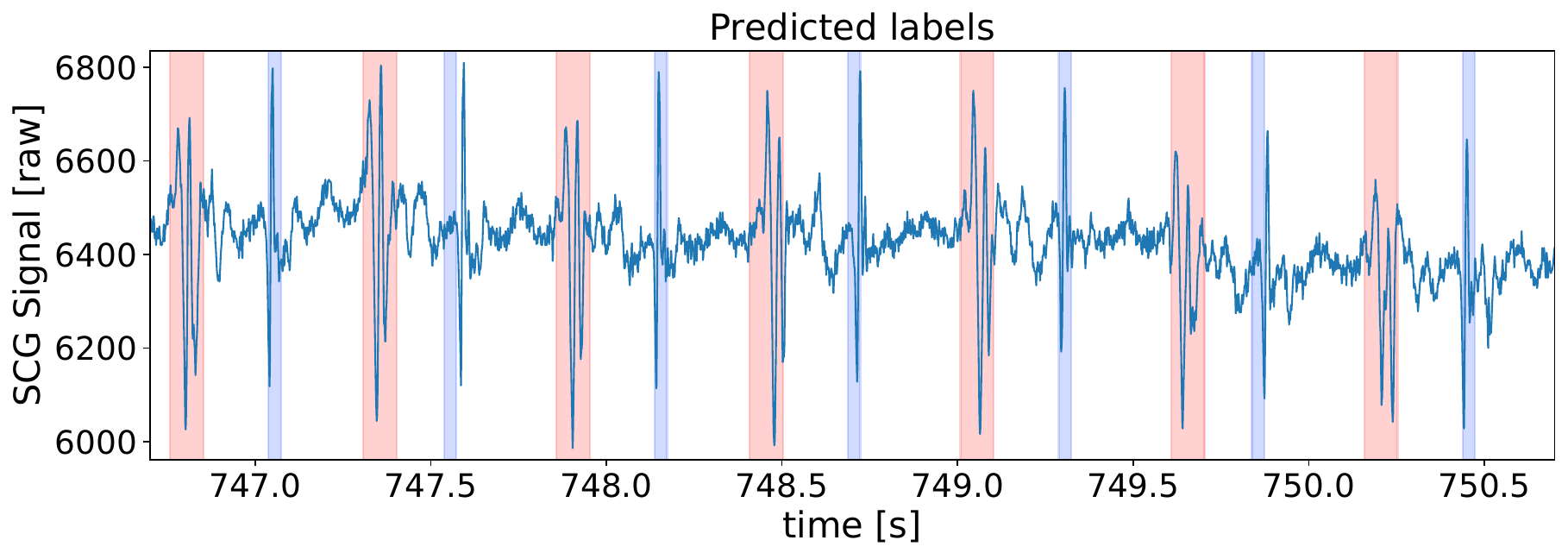}

  \caption{Comparison of ground-truth (top) and CNN-predicted (bottom) labels for the same SCG segment. Shaded regions indicate systolic (red) and diastolic (blue) intervals over the raw signal (small excerpt of 7 heartbeats).}
  \label{fig:gt_vs_pred_segment}
\end{figure}

\begin{figure}[t]
  \centering

  \begin{subfigure}[t]{0.48\linewidth}
    \centering
    \includegraphics[width=\linewidth]{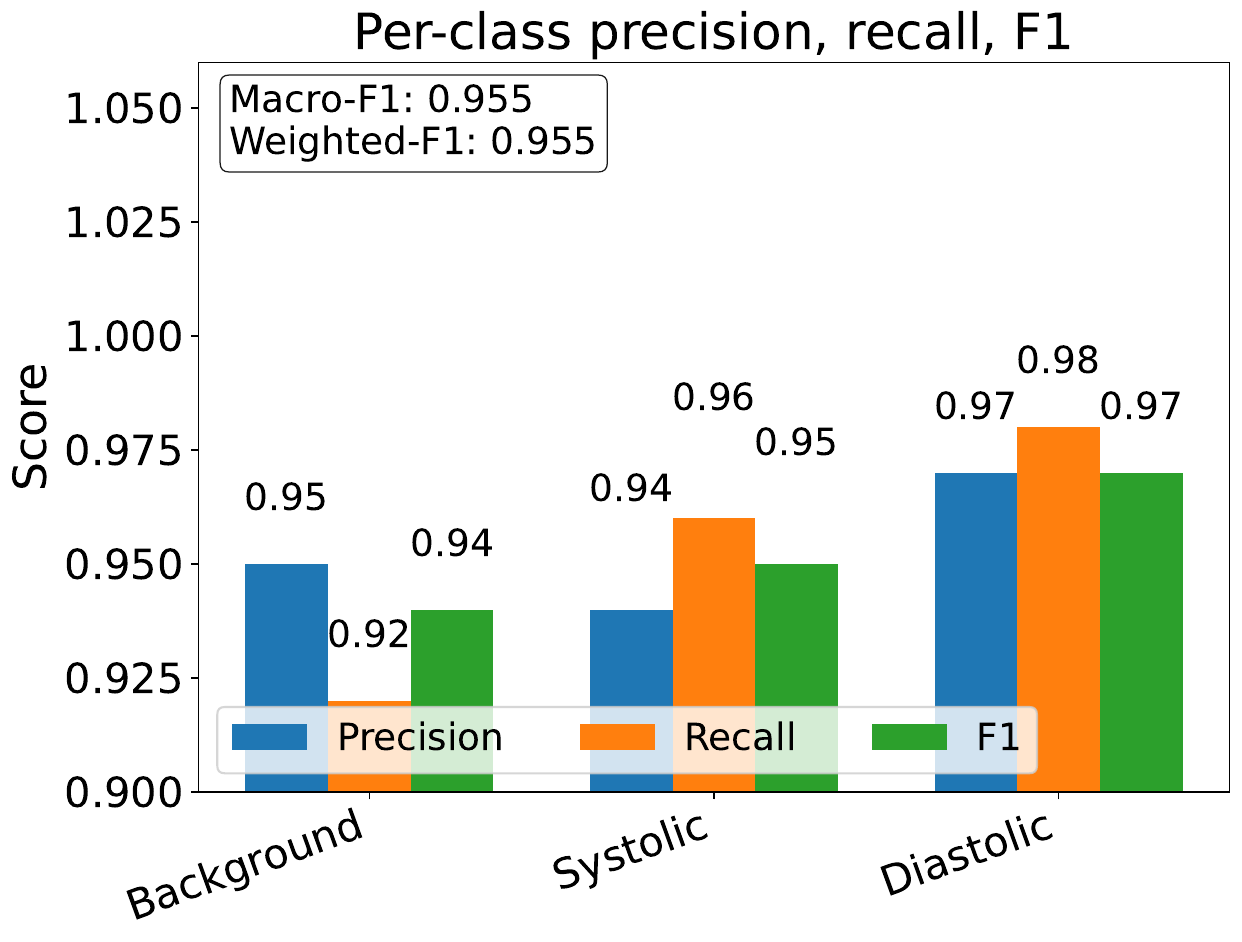}
    \caption{}
    \label{fig:eval_metrics}
  \end{subfigure}
  \hfill
  \begin{subfigure}[t]{0.48\linewidth}
    \centering
    \includegraphics[width=\linewidth]{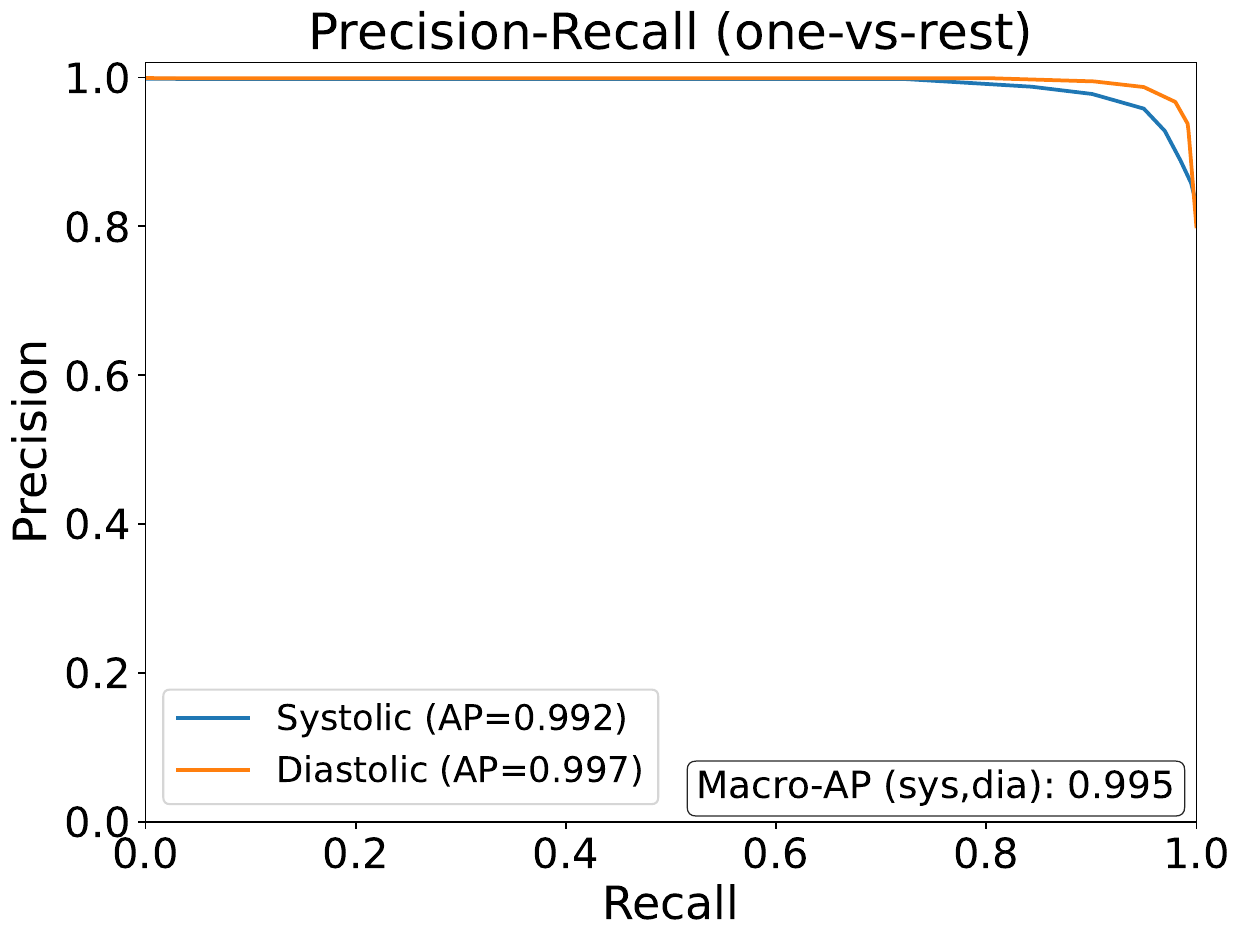}
    \caption{}
    \label{fig:eval_pr}
  \end{subfigure}

  \vspace{0.5em}

  \begin{subfigure}[t]{0.48\linewidth}
    \centering
    \includegraphics[width=\linewidth]{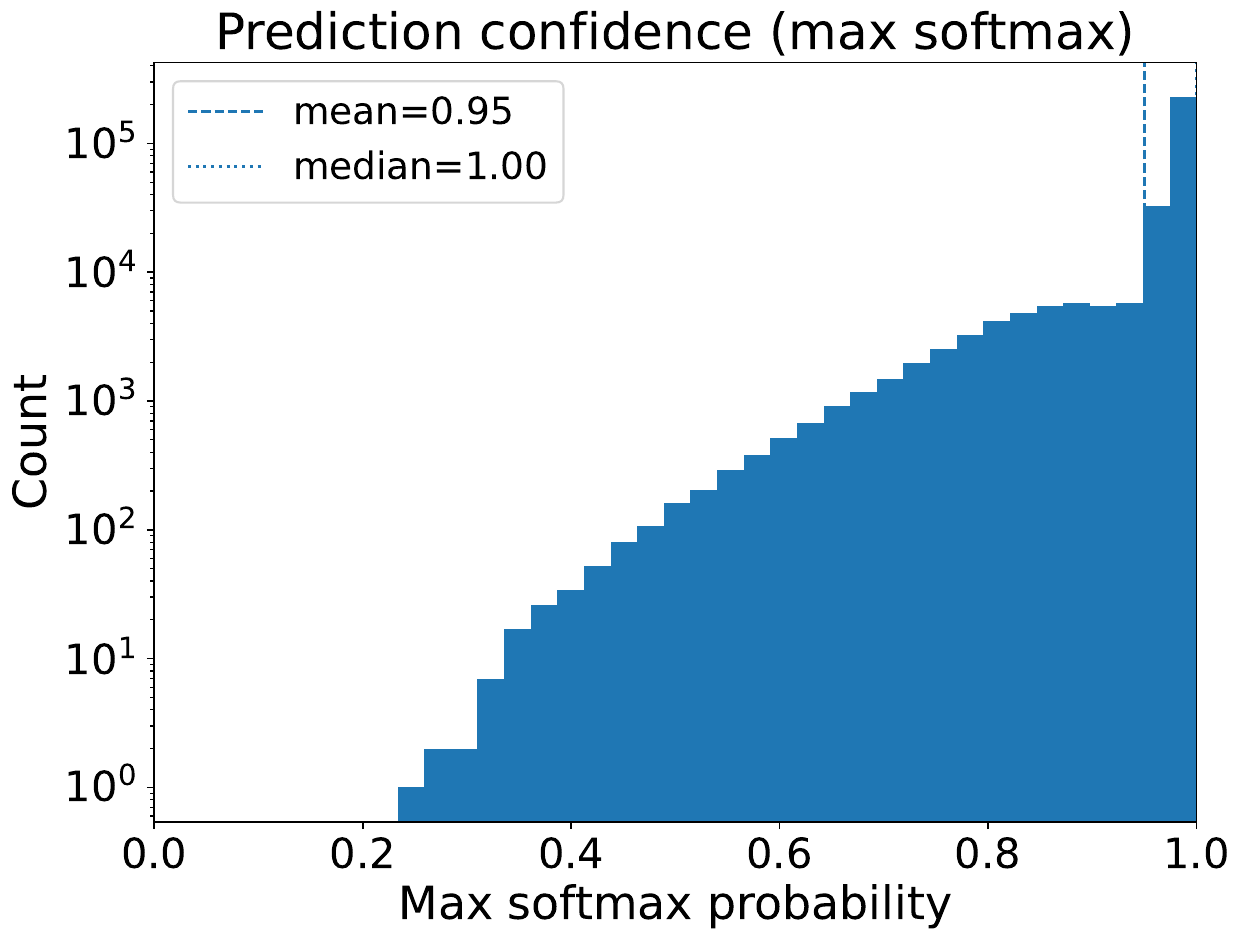}
    \caption{}
    \label{fig:eval_confidence}
  \end{subfigure}
  \hfill
  \begin{subfigure}[t]{0.48\linewidth}
    \centering
    \includegraphics[width=\linewidth]{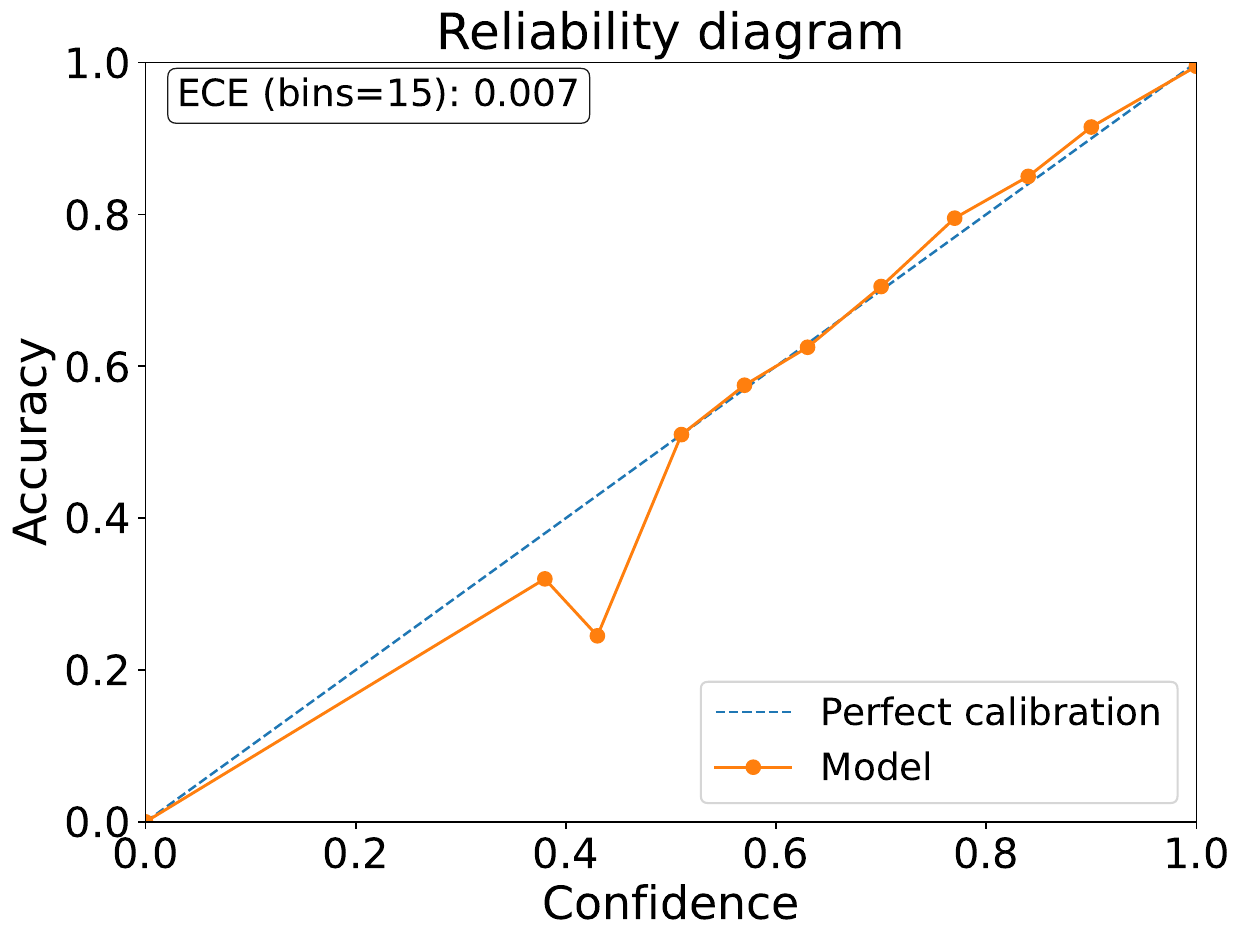}
    \caption{}
    \label{fig:eval_reliability}
  \end{subfigure}

  \caption{Combined window-level evaluation of the \ac{SCG} \ac{CNN}: (a) per-class precision, recall, and F1 with macro and weighted F1; (b) one-vs-rest precision-recall curves for systolic and diastolic events with average precision (AP); (c) histogram of maximum softmax confidence; (d) reliability diagram with \acf{ECE}.}
  \label{fig:eval_all_metrics}
\end{figure}

The proposed model demonstrates strong window-level discrimination for rare \ac{SCG} events.
Figure~\ref{fig:gt_vs_pred_segment} shows a qualitative comparison between ground-truth and \ac{CNN}-predicted labels on a representative \ac{SCG} segment, illustrating good temporal alignment of systolic and diastolic events.

As shown in Figure~\ref{fig:eval_all_metrics}, quantitative performance is maintained under severe class imbalance, with both systolic and diastolic events exhibiting high separability in precision--recall space.
Per-class metrics further confirm consistent behavior across all labels, indicating that performance is not dominated by the background class.

To reduce ambiguity near event boundaries, background windows were sampled using a temporal exclusion constraint, enforcing a minimum gap of \SI{0.05}{\s} from any annotated systolic or diastolic event.

\subsection{Inference Performance Evaluation}
\label{sec:inference_performance}

\begin{table}[t]
\parnotereset
\caption{iCE40UP5K Resource Utilization}
\label{tab:resource_util}
\centering
 \setlength{\tabcolsep}{2pt}
\begin{tabularx}{\linewidth}{p{1.5cm} c c c X}
\toprule
\textbf{Resource} &
\textbf{Used} &
\textbf{Total} &
\textbf{Util.} &
\textbf{Description} \\
\midrule
LUTs   & $\sim$2861 & 5280 & 54\%  & FSMs, address calculation, multiplexers \\
SPRAM\parnote{(256 kbit)}   & 4          & 4    & 100\% & Weight storage and feature maps         \\
BRAM\parnote{(4 kbit)} & 6          & 30   & 33\%  & Input buffers and bias ROM      \\
DSP Blocks         & 7          & 8    & 87\%  & Compute DSP Cluster and Requant logic      \\
IO Pins            & 14         & 96   & 14\%  & UART interface, debug bus, LEDs         \\
\bottomrule
\end{tabularx}
\parnotes
\end{table}

\begin{figure}
    \centering
    \includegraphics[width=1\linewidth]{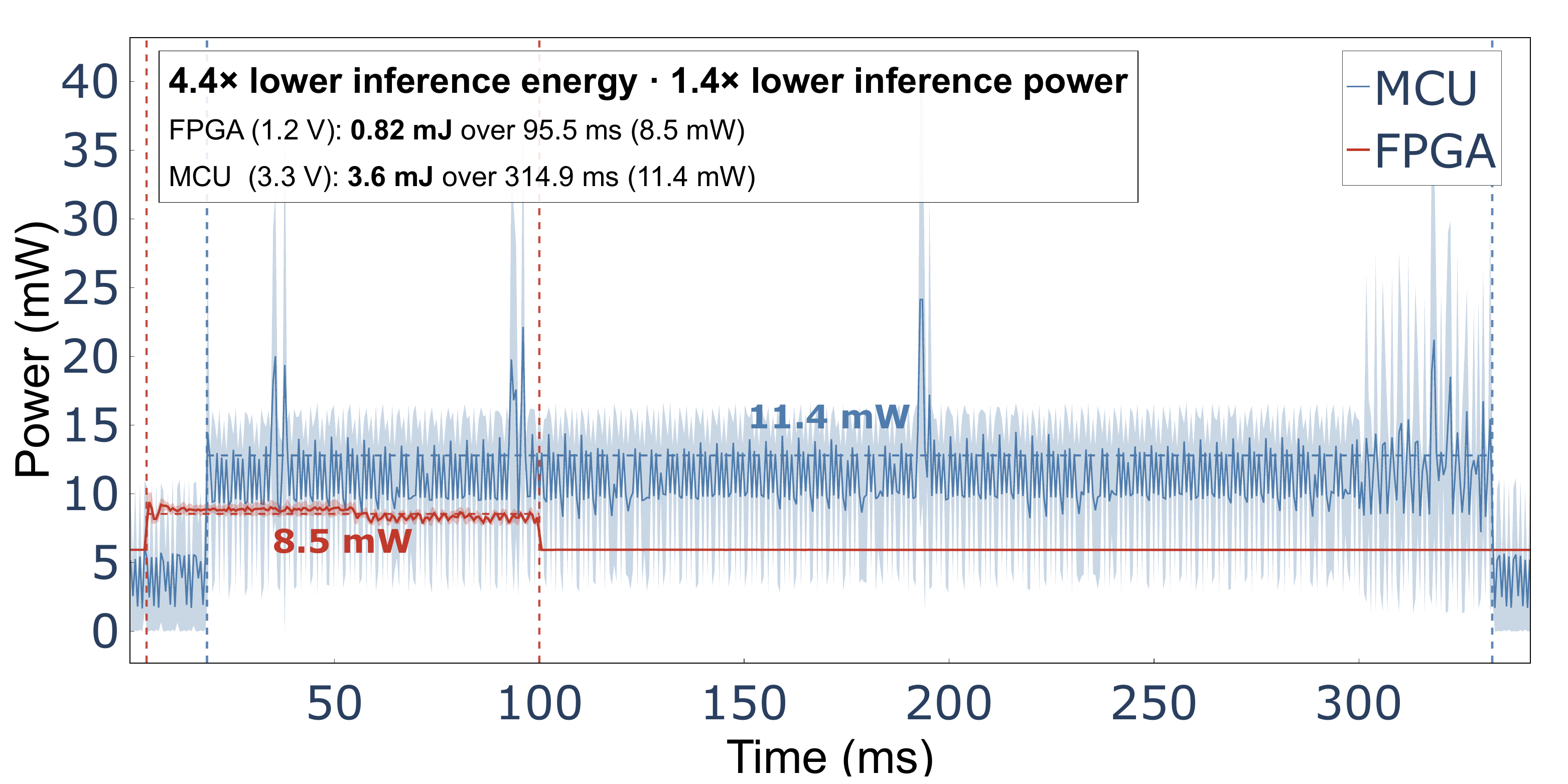}
    \caption{Inference Power Consumption comparison between iCE40UP5K FPGA and nRF52840 \ac{MCU}}
    \label{fig:infPower}
\end{figure}

\begin{table}[t]
\centering
\caption{Inference Efficiency and Performance Comparison Between \ac{FPGA} and \ac{MCU} Implementations}
\label{tab:comparison}
\setlength{\tabcolsep}{3pt}
\begin{tabularx}{\linewidth}{lcc}
\toprule
\textbf{Metric} & \textbf{iCE40UP5K (FPGA)} & \textbf{nRF52840 (MCU)} \\
\midrule
Inference Time [ms]            & 95.5   & 314.9 \\
Avg. Inference Power [mW]      & 8.55   & 11.4  \\
Inference Energy [$\mu$J]      & 819.1  & 3589.9 \\
Throughput [MMAC/s]            & 67.0   & --    \\
Throughput [MOps/s]            & 134.0  & --    \\
\bottomrule
\end{tabularx}
\end{table}

The \ac{FPGA} implementation demonstrates efficient logic utilization, requiring only 2,861 \acp{LUT} and 7 \acp{DSP}, as shown in Table~\ref{tab:resource_util}.
These results indicate that the neural network maps effectively onto the \ac{FPGA} fabric and preserves substantial capacity for additional features or future scaling.
In comparison to the energy-efficient nRF52840 \ac{MCU}, which operates at \SI{64}{\MHz} and uses CMSIS-NN to accelerate the LiteRT neural network inference with ARM \ac{DSP} cores, inference on the iCE40UP5K \ac{FPGA} (operating at \SI{24}{\MHz}) consumes only \SI{819}{\micro\joule} per inference.
This energy consumption is more than four times lower than that of the MCU, and inference completes in \SI{95.5}{\ms}, making it approximately three times faster (see Table~\ref{tab:comparison}).
The architecture achieves a throughput of 134~MOps/s while maintaining a low average power consumption of \SI{8.55}{\mW}, as shown in Figure~\ref{fig:infPower}.
These findings suggest that the architecture is suitable for ultra-low-power, resource-constrained \acp{FPGA} and is scalable to larger devices.

\section{Conclusion}
\label{sec:conclusion}

In this paper, we demonstrate a complete pipeline for real-time cardiac feature extraction (systolic-diastolic segmentation) for a space-grade \ac{SCG} wearable, encompassing data acquisition, high-fidelity semi-automated labeling, \ac{CNN} training, and energy-efficient inference on \ac{ULP} \acp{FPGA}.
Quantization-aware training enables \ac{CNN} models to achieve over \SI{98}{\percent} validation accuracy, while the systolic-array \ac{FPGA} implementation allows efficient on-device inference, completing classification in about \SI{95.5}{\ms} using only 2,861 \acp{LUT} and 7 \acp{DSP} ($\sim$10.6 FPS) at \SI{8.55}{\mW}, which is over four times more energy-efficient than state-of-the-art low-power \acp{MCU}.
Layerwise inference-cycle profiling identifies optimization opportunities in re-quantization and pipeline balance.
This full-cycle system is robust to sensor variability and enables fully autonomous, energy-efficient cardiovascular monitoring for astronauts, with broader implications for wearable and smart health technologies.

\section*{Acknowledgement}
This work has received funding from the  Federal Ministry of Research, Technology and Space (BMFTR) supported by German Aerospace Center (DLR) in the projects `AuRelia' and `SArES' under the grant numbers 50RP2350 and 50WB2421A, respectively. Also partially funded by the German Research Foundation (DFG) in the project `KORVEKSiS' under the grant number 542151450. The authors would like to express their sincere gratitude to Prof. Dr. Maximilian Kiener for facilitating the ethical approval of this study.

\bibliographystyle{IEEEtran}
\balance
\bibliography{literature}

\end{document}